\documentclass[longauth]{aa}  

\usepackage[switch]{lineno}
\usepackage{url}
\usepackage{graphicx}
\usepackage{xcolor}
\usepackage{natbib}
\bibpunct{(}{)}{;}{a}{}{,}
\usepackage{txfonts}
\usepackage{hyperref}
\begin{document} 
\linenumbers

   \title{Complex Organic Molecules in Protostars with ALMA Spectral Surveys (COMPASS)}
   \subtitle{V. Tracing cavity walls and shocked knots with nonthermally desorbed CH$_3$OH in BHR71-IRS1}

   \author{H.-S. Yun\inst{1,2}
          \and
          J.-E. Lee\inst{1,3}
          \and
          J. K. J{\o}rgensen\inst{4}
          \and
          A. Coutens\inst{5}
          \and
          A. L. Plunkett\inst{6}
          \and
          M. N. Drozdovskaya\inst{7,8}
          \and
          A. Belloche\inst{9}
          \and
          J.-H. Jeong\inst{1}
          \and
          P. Nazari\inst{10}
          \and
          Y.-L. Yang\inst{11}
          \and
          M. L. R. van 't Hoff\inst{12}
          \and
          S. Maret\inst{13}
          \and
          F. Cruz-S\'aenz de Miera\inst{5}
          \and
          D. Harsono\inst{14}
          \and
          C.-H. Kim\inst{1}
          \and
          \'A. K\'osp\'al\inst{15,16,17}
          \and
          M. L{\"u}tzen\inst{4}
          \and
          S. Spezzano\inst{18}
          \and
          S. Zeng\inst{11}
          }

   \institute{Department of Physics and Astronomy, Seoul National University, 1 Gwanak-ro, Gwanak-gu, Seoul 08826, Republic of Korea\\
              \email{yun.hyeongsik@snu.ac.kr; lee.jeongeun@snu.ac.kr}
         \and
              Korea Astronomy and Space Science Institute, 776 Daedeok-daero, Yuseong-gu, Daejeon 34055, Republic of Korea
         \and
              SNU Astronomy Research Center, Seoul National University, 1 Gwanak-ro, Gwanak-gu, Seoul 08826, Republic of Korea
        \and
             Niels Bohr Institute, University of Copenhagen, Jagtvej 155A, DK-2200 Copenhagen N, Denmark
        \and
             Institut de Recherche en Astrophysique et Plan\'etologie, Universit\'e de Toulouse, CNRS, CNES, 9 av. du Colonel Roche, 31028 Toulouse Cedex 4, France 
        \and 
             National Radio Astronomy Observatory, 520 Edgemont Road, Charlottesville, VA 22903, USA
        \and 
             Physikalisch-Meteorologisches Observatorium Davos und Weltstrahlungszentrum (PMOD/WRC), Dorfstrasse 33, 7260 Davos Dorf, Switzerland
        \and 
             Department of Chemistry, Biochemistry and Pharmaceutical Sciences (DCBP), Universit{\"a}t Bern, Freiestrasse 3, 3012 Bern, Switzerland
        \and 
             Max-Planck-Institut f{\"u}r Radioastronomie, Auf dem H{\"u}gel 69, 53121 Bonn, Germany
        \and 
             European Southern Observatory, Karl-Schwarzschild-Strasse 2, 85748 Garching, Germany
        \and 
             Star and Planet Formation Laboratory, RIKEN Pioneering Research Institute, Wako-shi, Saitama, 351-0198, Japan
        \and 
             Purdue University, 525 Northwestern Ave, West Lafayette, IN 47907 USA
        \and 
             Univ. Grenoble Alpes, CNRS, IPAG, 38000 Grenoble, France
        \and 
             Institute of Astronomy, Department of Physics, National Tsing Hua University, No. 101, Section 2, Kuang-Fu Road, Hsinchu 30013, Taiwan, R.O.C
        \and 
             Konkoly Observatory, HUN-REN Research Centre for Astronomy and Earth Sciences, MTA Centre of Excellence, Konkoly-Thege Mikl\'os \'ut 15-17, 1121 Budapest, Hungary
        \and 
             Institute of Physics and Astronomy, ELTE E\"otv\"os Lor\'and University, P\'azm\'any P\'eter s\'et\'any 1/A, 1117 Budapest, Hungary
        \and 
             Max-Planck-Institut f\"ur Astronomie, K\"onigstuhl 17, 69117 Heidelberg, Germany
        \and 
             Max-Planck-Institut f{\"u}r extraterrestrische Physik, Giessenbachstrasse 1, 85748 Garching bei M{\"u}nchen, Germany
             }

  \abstract
   {Complex organic molecules (COMs) are detected in the interstellar medium, both in cold molecular clouds and in the warm inner regions of protostars. Whether these COMs are eventually directly inherited by protostars and planets or whether their chemistry is reset by in situ reactions remains unclear. Recent (sub)millimeter wavelength observations have spatially resolved the emission of gas-phase COMs around low-mass protostars, not only in compact hot corinos but also in the extended structures beyond them. Understanding the physical properties and desorption mechanisms of these molecules is essential for tracing their chemical evolution during the protostellar evolutionary sequence. Methanol (CH$_3$OH), the most abundant COM in star-forming regions, serves as an ideal tracer for investigating these processes.}
   {As part of the ALMA Cycle 9 Large Program Complex Organic Molecules in Protostars with ALMA Spectral Surveys (COMPASS), we investigate the CH$_3$OH emission that extends beyond the hot corino of the low-mass protostar BHR71-IRS1.}
   {We identified spatially extended emission of CH$_3$OH around BHR71-IRS1 and subsequently explored plausible desorption scenarios. We assessed characteristic emission structures by applying a principal component analysis (PCA) to the identified emission lines. The physical properties of their line-emitting gas were derived under the local thermodynamic equilibrium (LTE) assumption through both rotational diagram and spectral modeling analyses.}
   {In total, 26 spatially extended CH$_3$OH lines have been detected around BHR71-IRS1. These emission lines are distributed across three distinct structures: (1) X-shaped cavity walls, (2) compact northern and southern shocked knots, and (3) additional distant eastern and southeastern components. The cavity walls exhibit excitation temperatures of $\sim$ 25~K and column densities of $\sim$ 10$^{14}$~cm$^{-2}$, while the knots show relatively higher excitation temperatures (70$-$120~K) and column densities ($10^{14}-10^{16}$~cm$^{-2}$). The eastern and southeastern components appear unrelated to BHR71-IRS1, and their physical properties are not well constrained under LTE conditions, suggesting that a non-LTE analysis may be required.}
   {Plausible desorption mechanisms are evaluated by comparing the physical parameters with model predictions that simulated the effects of far-ultraviolet (FUV) photons and C-type shocks. The low excitation temperatures and narrow line widths ($\sim$ 0.7~km~s$^{-1}$) in the cavity walls favor a reactive desorption scenario driven by FUV irradiation from IRS1. In contrast, the knots with higher excitation temperatures and column densities support the dust sputtering scenario induced by C-type shocks.}
   
   \maketitle
   \nolinenumbers

\section{Introduction}
Complex organic molecules (COMs), defined as carbon-bearing molecules composed of six or more atoms \citep{her09}, represent key building blocks of prebiotic chemistry. COMs have been detected both in dense starless and prestellar cores \citep{Taf06,Obe10,Bac12,Vast14,Sci20} and in the vicinity of low-mass protostars \citep{Cec05,Bis08,Kri10,LeeJE19,Man20}, and they may hint at the chemical inheritance from cold molecular clouds to planetary systems. In cold molecular clouds, COMs primarily form on icy dust grains through grain-surface chemistry \citep{Tie82,Ehr00,Chu18,Yan22}, and they are subsequently released into the gas phase via thermal sublimation or nonthermal desorption. Once desorbed, they undergo gas-phase reactions, photodissociation, freeze-out, or ionization \citep{Her86,Gar13,Dro15,Bur19,Cec23}, which can potentially alter their composition before incorporation into protoplanetary disks and planets \citep{Obe21}. Thus, to understand the chemical evolution linking molecular clouds to planet-forming environments, it is essential to characterize the desorption mechanisms and physical environments of gas-phase COMs.

Among COMs, methanol (CH$_3$OH) is both the simplest and most abundant, making it a robust tracer of gas-phase organics. Strong CH$_3$OH emission is commonly observed near protostars where dust temperatures exceed the $\sim$100 K sublimation threshold, producing compact hot-corino structures \citep[e.g., ][]{Caz03,Cec05,Cas12,Jor20}. However, CH$_3$OH emission has increasingly been detected on scales significantly larger than hot corinos \citep{Arc08,Kri10,Cod20,LeeSH24}, indicating the presence of nonthermal desorption. The physical origin of this extended component remains poorly understood, largely due to limited observational constraints on its spatial distribution and excitation conditions. Proposed mechanisms include photodesorption by far-ultraviolet (FUV) radiation, energetic-particle processing, reactive desorption, and dust sputtering in shocks \citep{Gar07,Bur19,Dro15,Chu18}. Distinguishing between these mechanisms requires high-resolution, multi-transition observations capable of separating extended components from compact hot corino emission.

The advent of the Atacama Large Millimeter/submillimeter Array (ALMA) has enabled spatially resolved COM studies with a high angular resolution and sensitivity. ALMA spectral surveys additionally provide broad molecular inventories and excitation constraints \citep{Jor20,LeeJE24}. To systematically investigate the COM chemistry in low-mass protostars, the Complex Organic Molecules in Protostars with ALMA Spectral Surveys\footnote{\url{https://erda.ku.dk/vgrid/COMPASS/}} (COMPASS) Large Program \citep[ALMA proposal id: 2022.1.00316.L; ][]{Jes25compass} obtained unbiased Band 7 spectral survey data toward 11 embedded protostellar regions across diverse environments and evolutionary stages. The achieved angular resolution isolates extended emission, enabling a detailed assessment of its spatial and excitation properties

Among the COMPASS sources, BHR71-IRS1 is uniquely well suited for studying extended CH$_3$OH emission due to its proximity \citep[176 pc; ][]{Jes25compass}, isolation, and well-characterized bipolar outflows \citep{Bou97,Par06,Ben17,Zap18}. The system hosts two Class 0 protostars separated by $\sim$2800 au \citep{Bou01}, with IRS1 being the brighter and more evolved member \citep[$L_\mathrm{bol}=9.2~L_\odot$, $T_\mathrm{bol}=49$~K; ][]{Jes25compass}, compared to IRS2 \citep[$L_\mathrm{bol}=1.7~L_\odot$ and $T_\mathrm{bol}=38$~K; ][]{Tob19}. Shocked gas associated with the bipolar outflows, extending $\sim$100\arcsec \ to the north and south of the protostars, has been identified in various molecular emission lines \citep[e.g., H$_2$, SiO, and high-J CO; ][]{Neu09,Gia11,Gus11}. Model comparisons revealed that this gas was affected by partially dissociative shocks with a velocity of $\sim$20~km~s$^{-1}$ approximately 4000 years ago \citep{Gus15,Ben17}.

Several nonthermal desorption mechanisms have been proposed to explain extended CH$_3$OH emission: (1) photodesorption by FUV photons, (2) desorption induced by energetic particles, (3) dust sputtering in shocks \citep{Bur19}, and (4) reactive desorption following grain-surface reactions \citep{Gar07,Dro15,Chu18}. Although photons and energetic particles often drive surface chemistry rather than desorb intact CH$_3$OH \citep{Ben07,deB11,Che13}, both reactive desorption and dust sputtering remain plausible channels for producing the observed extended COM emission.

For this study, we investigated the spatial distribution and excitation of CH$_3$OH toward BHR71-IRS1 and characterized the nature of its extended emission. We identified CH$_3$OH transitions with a spatially extended structure beyond the hot-corino region and performed local thermodynamic equilibrium (LTE) analysis to derive physical parameters. The paper is organized as follows: Sect.~\ref{sec:data} describes the COMPASS data, Sect.~\ref{sec:analysis} outlines line-identification and analysis methods, Sect.~\ref{sec:result} presents the emission morphology and excitation structure, Sect.~\ref{sec:discuss} discusses implications for desorption mechanisms, and Sect.~\ref{sec:summary} summarizes the conclusions.

\section{Observations and data reduction} \label{sec:data}
The COMPASS data were obtained in ALMA Band 7 during Cycle 9 and 10 (2022.1.00316.L). The spectral survey spans 279--312~GHz and is covered by nine spectral settings. Each setting comprises four spectral windows with 937~MHz bandwidth and 488.28~kHz spectral resolution, corresponding to velocity resolutions of 0.524~km~s$^{-1}$ at 279~GHz and 0.470~km~s$^{-1}$ at 311~GHz. The observations of BHR71-IRS1 were conducted from 8 to 22 January 2023. The motivation for the COMPASS program and its overall strategy can be found in \citet{Jes25compass}. For more details about the observations and data reduction, we refer the reader to \citet{Ade25compass}.  

We adopted the standard spectral cubes from the COMPASS pipeline \citep{Mar25compass}, which are continuum-subtracted. These data provide sufficient angular resolution and sensitivity to detect weak, spatially extended emission of COM lines. The rms noise level across all spectral windows ranges from 0.17 to 0.25~K, enabling the identification of very faint emission, with peak intensities as low as 1~K, at an adequate signal-to-noise ratio. The synthesized beam varies slightly across the observed frequency range, ranging from 0\farcs51 $\times$ 0\farcs38 to 0\farcs42 $\times$ 0\farcs33. Acquired spectral cubes generally achieve an angular resolution higher than 0\farcs5 ($<$ 90~au), allowing a detailed characterization of the gas distribution in the vicinity of IRS1. 

To enable direct comparisons among all detected CH$_3$OH lines, we convolved all data to a common beam determined with the ``commonbeam\_with" algorithm in the ``radio-beam" Python package\footnote{\url{https://radio-beam.readthedocs.io/en/latest/index.html}}. The resulting smallest common beam is 0\farcs52 $\times$ 0\farcs44 at a position angle of 34.87\degr. All cubes are also reprojected onto a common spatial grid.

\section{Analyses} \label{sec:analysis}
\subsection{Identification of a central component} \label{sec:anal_corino}
Near the protostar, numerous COM emissions arise not only from a compact hot corino \citep[with an emission size of 0\farcs7; ][]{Aud25compass} but also from extended line-emitting gas in the vicinity. To isolate emission lines that arise solely from the extended line-emitting gas, we first delineated the central component of BHR71-IRS1, following the method of \citet{Yun23}. From the beam-averaged spectrum extracted at the continuum peak at $(\alpha,\delta)_{\rm ICRS}=(12^{\mathrm h}01^{\mathrm m}36\fs475,,-65\degr08\arcmin49\farcs36)$, we identified COM transitions that exclusively trace the central component.

Because numerous weak COM transitions fall within the surveyed bandpass, transitions are frequently blended, often in ways that are difficult to recognize. Although such blending can be addressed with full forward spectral modeling, we instead searched for isolated COM transitions using cataloged spectroscopic parameters (e.g., $A_{\rm ij}$ and $E_\mathrm{up}$) for species commonly found in protostellar environments. To this end, we compiled spectroscopic data for both simple molecules and complex organic molecules detected in recent spectral surveys.

We incorporated the list of molecules identified in other COMPASS studies. \citet{Jes25compass} provides an overview of the COMPASS ALMA Large Program and presents the full hot corino spectrum of BHR71-IRS1, highlighting strong lines from simple molecules. Several complementary papers analyze various COMs and their isotopologs to derive physical parameters: methanol \citep{Aud25compass}, methyl cyanide \citep{Poo25compass}, and sulfur-bearing species (Harsono et al. in prep.). We also included COMs reported in other spectral surveys, namely the ALMA Protostellar Interferometric Line Survey \citep[PILS;][]{Cou16,Jor16,Fay17,Cal18b,Cal18a,Dro18,Jor18,Lig18,Mur18,Cal19,Man20,Man21,Dro22} and the ALMA Spectral Survey of An eruptive Young star, V883~Ori \citep[ASSAY;][]{LeeJE24,JeoJH25}. 

To facilitate the search for isolated COM lines, we compiled a comprehensive list of spectroscopic data for all considered molecules and isotopologs. The majority of the data were acquired from the Cologne Database for Molecular Spectroscopy \citep[CDMS; ][]{Mul01, Mul05, End16}\footnote{\url{https://cdms.astro.uni-koeln.de/cdms/portal/}}. Molecules not found in CDMS were obtained from the Jet Propulsion Laboratory database \citep{Pic98}\footnote{\url{https://spec.jpl.nasa.gov/home.html}} and the Lille Spectroscopic Database \citep[LSD; ][]{Mot25}\footnote{\url{https://lsd.univ-lille.fr/}}. The spectroscopic data for a total of 167 molecules and isotopologs were investigated, resulting in the inclusion of 54,631 transitions in our initial search. Since our prior search revealed that the full width at half maximum of emission lines is about 3.0~km~s$^{-1}$, we selected lines that satisfy: (1) peak line intensity $>1.0$ K, (2) upper state energy $E_\mathrm{up}<200$ K, (3) Einstein coefficient $A_{\rm ij}>10^{-5}\ \mathrm{s}^{-1}$, and (4) no evidence for any other emission line within $\pm5.0$ km s$^{-1}$. Five lines meet these criteria (Table~\ref{tbl:isol_str}; see also Fig. B.1; available on Zenodo\footnote{\url{https://doi.org/10.5281/zenodo.19499024}}). 

\begin{table}
\caption{COM lines for the identification of the central component.}
\label{tbl:isol_str}
\centering
\begin{tabular}{l c c c c c c}
\hline\hline
Species &\hspace*{-1ex}Frequency &\hspace*{-1ex}$I_c$\tablefootmark{a} &\hspace*{-1ex}$E_\mathrm{up}$ &\hspace*{-1ex}$\log A_{\rm ij}$ &\hspace*{-1.5ex}$g_\mathrm{up}$ &\hspace*{-1.5ex}Database\\
       & \hspace*{-1ex}(GHz) & \hspace*{-1ex}(K) & \hspace*{-1ex}(K) & \hspace*{-1ex}log(s$^{-1}$) & \hspace*{-1.5ex}& \\
\hline
CHD$_2$OH &\hspace*{-1ex}292.16727 &\hspace*{-1ex}2.57 &\hspace*{-1ex}68.5 &\hspace*{-1ex}-3.993 &\hspace*{-1.5ex}15 &\hspace*{-1.5ex}CDMS\\
CHD$_2$OH &\hspace*{-1ex}296.46062 &\hspace*{-1ex}1.97 &\hspace*{-1ex}107.5 &\hspace*{-1ex}-4.144 &\hspace*{-1.5ex}19 &\hspace*{-1.5ex}CDMS\\
CH$_2$DOH &\hspace*{-1ex}296.99540 &\hspace*{-1ex}3.09 &\hspace*{-1ex}58.4 &\hspace*{-1ex}-3.977 &\hspace*{-1.5ex}11 &\hspace*{-1.5ex}JPL\\
CH$_3$CHO &\hspace*{-1ex}297.30599 &\hspace*{-1ex}2.75 &\hspace*{-1ex}123.2 &\hspace*{-1ex}-3.029 &\hspace*{-1.5ex}33 &\hspace*{-1.5ex}LSD\\
$^{13}$CH$_3$CN &\hspace*{-1ex}303.60997 &\hspace*{-1ex}1.22 &\hspace*{-1ex}194.3 &\hspace*{-1ex}-2.627 &\hspace*{-1.5ex}14 &\hspace*{-1.5ex}CDMS\\
\hline
\end{tabular}
\tablefoot{
\tablefoottext{a}{The line intensities measured from the beam-averaged spectra extracted from the continuum peak position.}
}
\end{table}

To assess the distribution of the five lines selected above, we used principal component analysis (PCA), which provides an efficient way to characterize line emission distributions \citep{Ung97,Oko21,Yun23}. By decomposing the data into orthogonal principal components, PCA identifies the first principal component (PC1) that represents the overall line-emitting gas and separates distinct regions where their physical or chemical conditions may differ (higher-order PCs). This approach therefore provides a robust basis for subsequent detailed analyses. Since we already identified emission lines exclusively emitted from the central component, applying a three-dimensional principal component analysis (PCA3D) yields a meaningful single element (PC1) that represents the central component (see Appendix~\ref{App:PCA2D}). 

\begin{figure}
\centering
\includegraphics[width=0.5\textwidth]{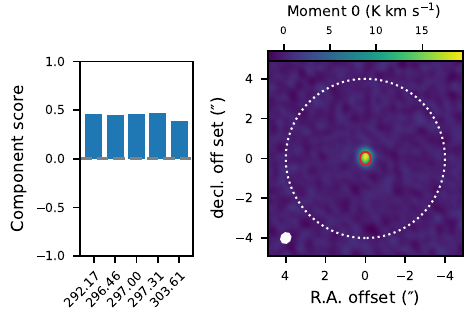}
\caption{Left: component scores of PC1 ($\vec{u}_1$). Each score is labeled with the rest frequency of the corresponding emission line listed in Table \ref{tbl:isol_str}. All component scores have the same positive sign and similar magnitudes, indicating a strong positive correlation among the five isolated COM lines. Right: integrated intensity map of the PC1-weighted cube, showing the representative emission structure of the central component. The integrated intensities are derived over $\pm$ 7.5~km~s$^{-1}$ from the system velocity of $-$4.45~km~s$^{-1}$. The white ellipse in the bottom left corner represents the smoothed beam size (0\farcs52 $\times$ 0\farcs44). The solid red line indicates the 2D Gaussian fitting result for the integrated intensity map (0\farcs67 $\times$ 0\farcs54), and the dotted white line indicates a circular aperture with a radius of 4\arcsec.} 
\label{fig:corino_mom0}
\end{figure}

The PC1 of PCA3D provides component scores (elements of the first eigenvector) that, when used as weights to combine the original cube, maximize correlated signal while suppressing noise (see Fig.~\ref{fig:corino_mom0}). Methodological details are given in \citet{Oko21}. For the selected COM lines, the PC1 component scores are all positive and of similar magnitude, indicating strong spatial and kinematic correlation and a common origin. The moment 0 map (integrated intensity map) of the PC1-weighted cube, derived from a $\pm$ 7.5~km~s$^{-1}$ regime from the system velocity of IRS1 ($-$4.45~km~s$^{-1}$), shows compact emission centered on IRS1. A two-dimensional (2D) Gaussian fit of the moment 0 map yields the size of the central component to be 0\farcs67 $\times$ 0\farcs54. Note that this estimated size does not represent the actual size of the hot corino since we are investigating the COM lines that have low $E_\mathrm{up}$ and high $A_\mathrm{ij}$.

\subsection{Identification of spatially extended molecular emission} \label{sec:identify} 
To achieve an unbiased search for lines with extended emission, we compared integrated intensities measured from spatially-averaged spectra extracted using two apertures: (1) a circular aperture of radius 4$\arcsec$ centered on the continuum peak, and (2) an aperture matching the measured size of the central component. Integrated intensities were subsequently derived for all available channels; for each spectral channel, integration is performed over a velocity range of $\pm$5 km s$^{-1}$, which encompasses all emission from the central component. 

For most COM lines, emissions would be concentrated within the central component. Therefore, the integrated intensity from the larger aperture ($I_\mathrm{extend}$), which is marked by a dashed white circle in the right panel of Fig. \ref{fig:corino_mom0} is lower than that from the aperture that matches the central component ($I_\mathrm{center}$), owing to averaging over regions with little or no emission. For a channel where an emission line exists and arises exclusively from the central component, the ratio $f_\mathrm{e/c} = I_\mathrm{extend}/I_\mathrm{center}$ should be equal to that derived from the representative structure of the central component (presented as a red ellipse in Fig.~\ref{fig:corino_mom0}), which we found to be $f_\mathrm{e/c} = 0.01257$. We can thus derive the expected intensity in the larger aperture for all channels, assuming the emission is confined to the central component, as
\begin{equation}
I_\mathrm{expect} = 0.01257 \times I_\mathrm{center}. \label{eqn:exp}
\end{equation}
A channel with $I_\mathrm{extend}$ significantly exceeding $I_\mathrm{expect}$ indicates the detection of extended emission. 

By assessing $I_\mathrm{extend}$ and $I_\mathrm{expect}$ for channels (frequencies) containing known CH$_3$OH transitions, we identified candidates for extended CH$_3$OH emission. Some of these candidates, however, exhibit elevated $I_\mathrm{extend}$ values because of blending with nearby extended CH$_3$OH transitions or with lines of simple molecules (e.g., H$_2$CO, CS, SO$_2$) located near the CH$_3$OH rest frequencies. To distinguish genuine extended emission, we inspected spectra at multiple positions, comparing line profiles and centroid velocities. After excluding blended or kinematically inconsistent features, we identify genuine extended CH$_3$OH lines.

This approach provides a first-order identification that is most effective for lines with strong, spatially extended emission. To supplement this, we also searched for extended CH$_3$OH transitions through visual inspection of spectra from representative positions during the spectral-fitting analysis (see Sect.~\ref{sec:modeling}). 

\subsection{Rotation diagram analysis and spectral fitting} \label{sec:modeling}
After identifying the CH$_3$OH lines with extended emission, we derived the physical properties of the emitting gas using the rotation (excitation) diagram method \citep{Gol99}. This approach provides first-order estimates of the rotational temperature ($T_\mathrm{rot}$) and the CH$_3$OH column density ($N_\mathrm{CH_3OH}$) under three standard assumptions: (1) all transitions are optically thin; (2) level populations are thermalized at a single excitation temperature; and (3) the emitting region is fully resolved (beam filling factor $\approx1$). For individual pixels with at least two detected CH$_3$OH lines, we performed rotation-diagram fits to produce maps of $T_\mathrm{rot}$ and $N_\mathrm{CH_3OH}$.

In every pixel, integrated intensities ($I_\mathrm{CH_3OH}$) and their uncertainties ($\sigma_\mathrm{CH_3OH}$) were measured with the moment-masking technique \citep{Dam11}, which identifies spectral channels and spatial pixels where emission exists. With this method, a velocity range containing at least one extended CH$_3$OH line is identified in each pixel. If the peak intensity of a certain line exceeded $3\sigma$, $I_\mathrm{CH_3OH}$ was obtained by integrating over the identified range; otherwise, a $3\sigma$ upper limit was adopted.

When three or more lines were detected above $3\sigma$, we estimated best-fit parameters via a bootstrap that incorporates upper limits:
\begin{enumerate}
\item Generate artificial measurements: $I_\mathrm{art}=I_\mathrm{CH_3OH}+\sigma_\mathrm{art}$, with $\sigma_\mathrm{art}$ drawn from a zero-mean Gaussian of width $\sigma_\mathrm{CH_3OH}$.
\item If $I_\mathrm{art}<0$, treat the line as undetected and substitute its $3\sigma$ upper limit.
\item For nondetections, draw artificial values below their upper limits.
\item Construct the rotation diagram from the artificial set and perform a least-squares linear fit; accept the fit only if all upper-limit constraints are satisfied.
\item Repeat until 100 valid fits are obtained; if fewer than 20 valid fits are found after $10^6$ trials, flag the result as unreliable.
\end{enumerate}
For two-line cases, we performed a simple linear fit, but likewise classified the results as unreliable.

Because rotation diagrams can bias parameters when assumptions are violated (e.g., optical depth effects or unresolved emission), we also double checked the derived parameters by conducting complementary LTE spectral modeling at selected positions using CASSIS\footnote{\url{https://cassis.irap.omp.eu}} \citep{Vas15}. In this analysis, we additionally identified CH$_3$OH lines not captured in Sect.~\ref{sec:identify}. Synthetic spectra were computed with spectroscopic data from the CDMS database. The LTE model includes five free parameters—excitation temperature ($T_\mathrm{ex}$), molecular column density ($N_\mathrm{CH_3OH}$), full width at half maximum (FWHM), line center velocity ($v_\mathrm{LSR}$), and emission size—and is fitted in two steps: (1) a coarse grid search over broad parameter ranges to locate the vicinity of the minimum, followed by (2) a Markov chain Monte Carlo exploration for a fixed emission size. We first adopted a fully resolved case (emission size $=10\arcsec$), then probed parameter degeneracies by repeating the procedure over a range of emission sizes.

\section{Results} \label{sec:result}
\subsection{Central component} \label{sec:corino}
\begin{figure*}
\centering
\includegraphics[width=0.85\textwidth]{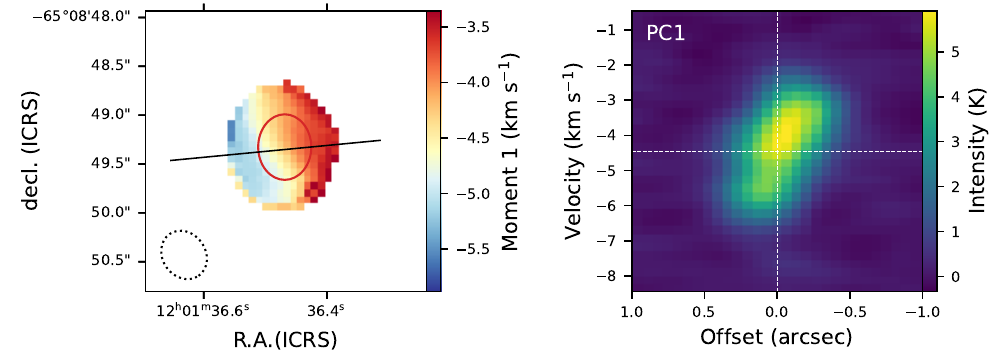}
\caption{Left: intensity-weighted mean velocity (moment 1) map of the central component obtained from the PC1-weighted cube. The red ellipse shows the 2D Gaussian fitting result of the integrated intensity map (0\farcs67 $\times$ 0\farcs54; Fig.~\ref{fig:corino_mom0}), and the dotted ellipse in the bottom left corner indicates the synthesized beam. The solid black line shows the spatial cut, perpendicular to the outflow axis, used to construct the position–velocity (PV) diagram. Right: PV diagram of the central component, extracted along the cut shown in the left panel.} 
\label{fig:PC1_pv}
\end{figure*}

Using five isolated COM lines (Table \ref{tbl:isol_str}), we constructed a PC1-weighted representative emission structure for the central component. Even though the central component is not well resolved with our beam size, it still allows us to measure its velocity gradient. Figure~\ref{fig:PC1_pv} presents the resulting velocity field; the intensity-weighted mean velocity (moment 1) map shows a modest gradient along the southeast–northwest direction, consistent with \citet{Tob19} and \citet{Yan20}. We extracted a position–velocity (PV) diagram along a $2\arcsec$ cut centered on the continuum peak and oriented perpendicular to the outflow axis (PA$=95.9^\circ$; \citealt{Gav24}). The PV structure matches that seen in $^{13}$CH$_3$OH by \citet{Yan20}. The presence of a rotating disk has not been confirmed yet, but these results imply the presence of gas-phase COMs around the outflow axis \citep{Aud25compass}.

\subsection{Methanol extended emission} \label{sec:extend}
\subsubsection{Lines showing extended emission}

\begin{figure}
\centering
\includegraphics[width=0.5\textwidth]{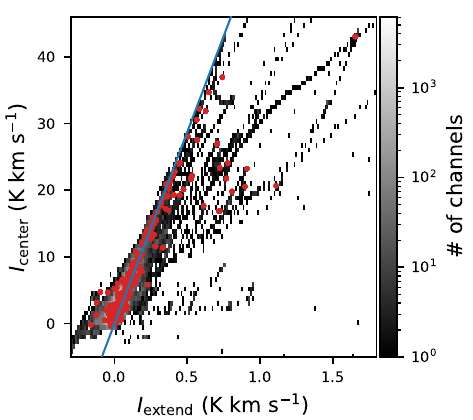}
\caption{Comparison between integrated intensities measured in the $4\arcsec$ aperture ($I_\mathrm{extend}$) and in the aperture which fits the central component ($I_\mathrm{center}$). The background density map shows the distribution of spectral channels in the $I_\mathrm{extend}$–$I_\mathrm{center}$ plane. The expected relation for emission arising solely from the central component (Eq.~\ref{eqn:exp}) is shown as the solid blue line. Red points mark channels that coincide with CH$_3$OH line frequencies.}
\label{fig:moving_integ}
\end{figure}

Figure~\ref{fig:moving_integ} shows the distribution of all velocity channels in the $I_\mathrm{extend}$–$I_\mathrm{center}$ plane. For $I_\mathrm{center}<10$ K km s$^{-1}$, the points follow the expected relation (solid blue line; $I_\mathrm{expect}=0.01257\times I_\mathrm{center}$), indicating no significant emission outside the central component. At higher values, $I_\mathrm{center}>10$ K km s$^{-1}$, $I_\mathrm{extend}$ increasingly exceeds $I_\mathrm{expect}$, implying additional extended emission.

We assessed the channels corresponding to the rest frequencies of all CH$_3$OH lines (the red circles in Fig.~\ref{fig:moving_integ}). Many CH$_3$OH lines show $I_\mathrm{extend}$ exceeding $I_\mathrm{expect}$ (lying substantially to the right of the blue line). The subsequent visual inspection of their line profiles yields thirteen CH$_3$OH lines with extended emission beyond the central component (typically characterized by $I_\mathrm{extend}-I_\mathrm{expect} \gtrsim 0.1$~K~km~s$^{-1}$). We also searched for additional extended CH$_3$OH emission during the spectral-fitting analyses (Sect.~\ref{sec:modeling}). This approach allows us to identify two additional isolated transitions, with small $I_\mathrm{extend}-I_\mathrm{expect}$ values, as well as four spectral features ($F_\mathrm{i}$) consisting of eleven overlapping CH$_3$OH transitions that were excluded during the prior visual inspection. In total, we identify 26 CH$_3$OH lines with extended emission (see Table \ref{tbl:extended}).

\begin{table*}
\caption{CH$_3$OH lines showing spatially extended emission.}
\label{tbl:extended}
\centering
\begin{tabular}{lllrrrrc}
\hline\hline
  \#\tablefootmark{a} & Transition & Frequency & $I_\mathrm{extend} - I_\mathrm{expect}$ & $E_\mathrm{up}$ & $\log A_\mathrm{ij}$ & $g_\mathrm{up}$ & $I$/$F_\mathrm{i}$\tablefootmark{b}\\
  & & (GHz) & (K~km~s$^{-1}$) & (K) & log(s$^{-1}$) & &\\
\hline
1&	3(1,2)$-$3(0,3) A, $v_t=0$&    305.473491&	0.519&	28.5&	-3.4863&  28& $I$\\
2&	2(1,1)$-$2(0,2) A, $v_t=0$&    304.208348&	0.443&	21.5&	-3.4903&  20& $I$\\
3&	4(1,3)$-$4(0,4) A, $v_t=0$&    307.165924&	0.483&	37.8&	-3.4807&  36& $I$\\
4&	6(0,6)$-$5(0,5) A, $v_t=0$&    290.110637&	0.368&	48.5&	-3.9745&  52& $I$\\
5&	1(1,0)$-$1(0,1) A, $v_t=0$&    303.366921&	0.288&	16.9&	-3.4931&  12& $I$\\
6&	5(1,4)$-$5(0,5) A, $v_t=0$&    309.290360&	0.340&	49.4&	-3.4738&  44& $I$\\
7&	6(1,6)$-$5(1,5) E, $v_t=0$&    290.069747&	0.293&	54.0&	-3.9867&  52& $I$\\
8&	6(-0,6)$-$5(-0,5) E, $v_t=0$&  289.939377&	0.124&	61.5&	-3.9757&  52& $I$\\
9&	3(-1,2)$-$2(-0,2) E, $v_t=0$&  310.192994&	0.059&	34.8&	-4.0555&  28& $I$\\
10&	6(-1,5)$-$5(-1,4) E, $v_t=0$&  290.248685&	0.060&	69.4&	-3.9753&  52& $I$\\
11&	6(1,5)$-$5(1,4) A, $v_t=0$&    292.672889&	0.104&	63.4&	-3.9748&  52& $I$\\
12&	6(1,6)$-$5(1,5) A, $v_t=0$&    287.670767&	0.125&	62.5&	-3.9973&  52& $I$\\
13\tablefootmark{c}& 12(0,12)$-$11(1,11) A, $v_t=0$& 302.912979& 0.053& 179.8& -3.8153& 100& $I$\\
14&	3(-0,3)$-$2(1,2) E, $v_t=0$&   302.369773&	0.094&	27.0&	-4.3315&  28& $I$\\
15&	6(2,4)$-$5(2,3) A, $v_t=0$&    290.264068&  0.013&    86.0&	-4.0222&  52& $I$\\ \hline
16&	6(4,3)$-$5(4,2) A, $v_t=0$&    290.161348&  0.003&   128.3&	-4.2288&  52& $F_1$\\
17&	6(4,2)$-$5(4,1) A, $v_t=0$&    290.161352&  0.003&   128.3&	-4.2288&  52& $F_1$\\
18&	6(4,3)$-$5(4,2) E, $v_t=0$&    290.162356&  0.006&   135.8&	-4.2296&  52& $F_1$\\
19&	6(-4,2)$-$5(-4,1) E, $v_t=0$&  290.183289& -0.012&   143.8&	-4.2269&  52& $F_2$\\
20&	6(2,5)$-$5(2,4) A, $v_t=0$&    290.184674&  0.009&    86.0&	-4.0226&  52& $F_2$\\
21&	6(3,4)$-$5(3,3) A, $v_t=0$&    290.189515&  0.061&    98.0&	-4.0997&  52& $F_2$\\
22&	6(3,3)$-$5(3,2) A, $v_t=0$&    290.190549&  0.040&    98.0&	-4.0997&  52& $F_2$\\
23&	6(3,3)$-$5(3,2) E, $v_t=0$&    290.209695&  0.002&   110.8&	-4.0966&  52& $F_3$\\
24&	6(-3,4)$-$5(-3,3) E, $v_t=0$&  290.213180& -0.012&    95.9&	-4.0984&  52& $F_3$\\
25&	6(2,4)$-$5(2,3) E, $v_t=0$&    290.307281&  0.210&    74.2&	-4.0242&  52& $F_4$\\
26&	6(-2,5)$-$5(-2,4) E, $v_t=0$&  290.307738&  0.209&    70.6&	-4.0294&  52& $F_4$\\
\hline
\end{tabular}
\tablefoot{The spectroscopic parameters are obtained from the CDMS database.
\tablefoottext{a}{The isolated lines are arranged in decreasing order of their component scores of PC2 (derived from the PCA analysis; see Sect.~\ref{sec:PCA2D}), which is an indicators of their spatial extents. The lines constituting the overlapped spectral features are arranged in increasing order of their central frequencies.}
\tablefoottext{b}{$I$ : isolated; $F_\mathrm{i}$: overlapping CH$_3$OH transitions constituting the spectral features $F_1$, $F_2$, $F_3$, and $F_4$.}
\tablefoottext{c}{Identified as a maser transition by \citet{Jeo25compass}.}
}
\end{table*}

\begin{figure*}
\centering
\includegraphics[width=0.9\textwidth]{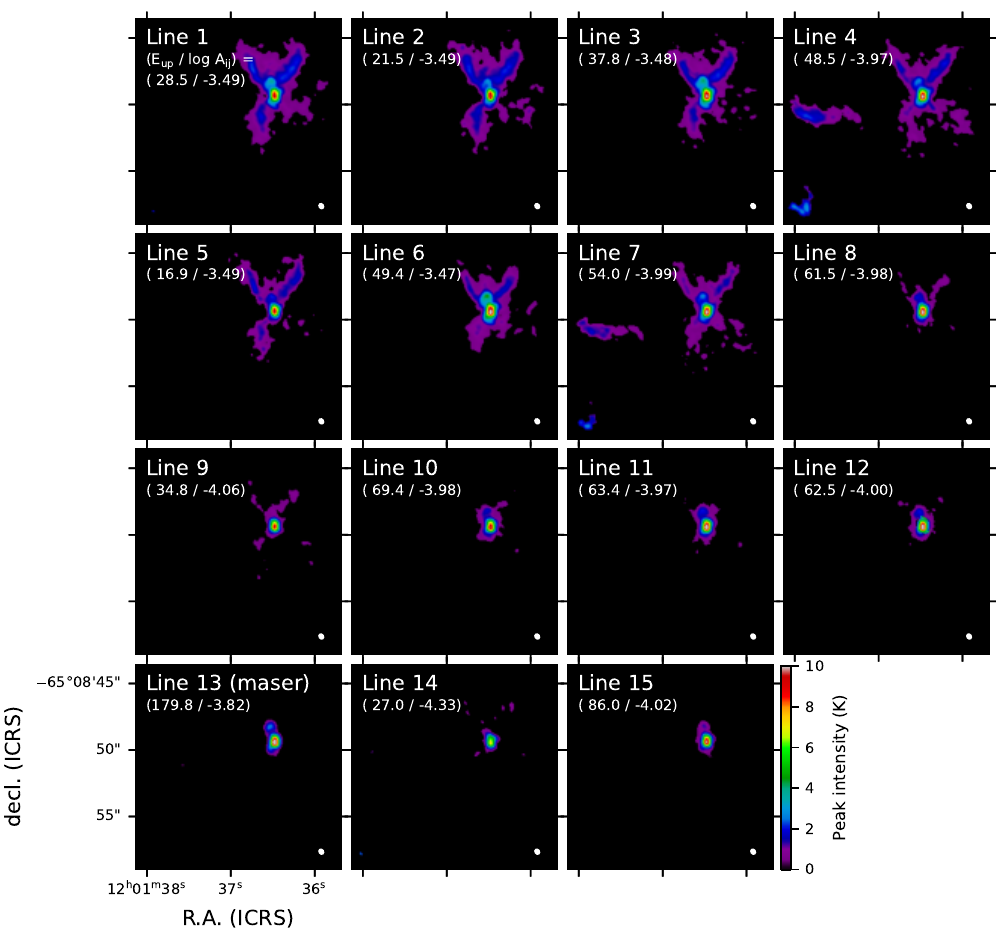}
\caption{Peak intensity maps of the spatially extended, isolated CH$_3$OH lines. Only pixels with the peak intensity values exceeding the $4\sigma$ noise level (around 0.5~K) are shown; other regions are blanked. The white ellipse in the bottom right corner of each panel indicates the synthesized beam (0\farcs52 $\times$ 0\farcs44). The $E_\mathrm{up}$ (K) and $\log (A_\mathrm{ij}\ (\mathrm{s}^{-1}))$ are listed in the top left corner of each panel.} 
\label{fig:mom8}
\end{figure*}

Figure~\ref{fig:mom8} displays the peak intensity maps of the fifteen isolated CH$_3$OH lines, illustrating the spatial extent of the line-emitting gas. The maps of the overlapped CH$_3$OH lines are provided as supplementary material on Zenodo (Fig.~C.1). The isolated lines were utilized for PCA to derive their spatial distribution and for rotation diagram analysis to derive their physical properties. The CASSIS modeling can also handle line blending. Therefore, all 26 lines were included in the CASSIS fitting analysis.

\subsubsection{Spatial structures traced by the methanol lines} \label{sec:PCA2D}
We assessed the spatial extent of the detected CH$_3$OH lines using PCA. Specifically, we computed two-dimensional PCA \citep[PCA2D;][]{Oko21} on the peak intensity maps of the 15 isolated CH$_3$OH transitions (Fig.~\ref{fig:mom8}) to reveal correlations among lines. We focused on the first four principal components (PC1 to PC4), which are physically meaningful among the decomposed PC$i$; methodological details and the selection of significant components are provided in Appendix~\ref{App:PCA2D}.

\begin{figure*}
\centering
\includegraphics[width=1.0\textwidth]{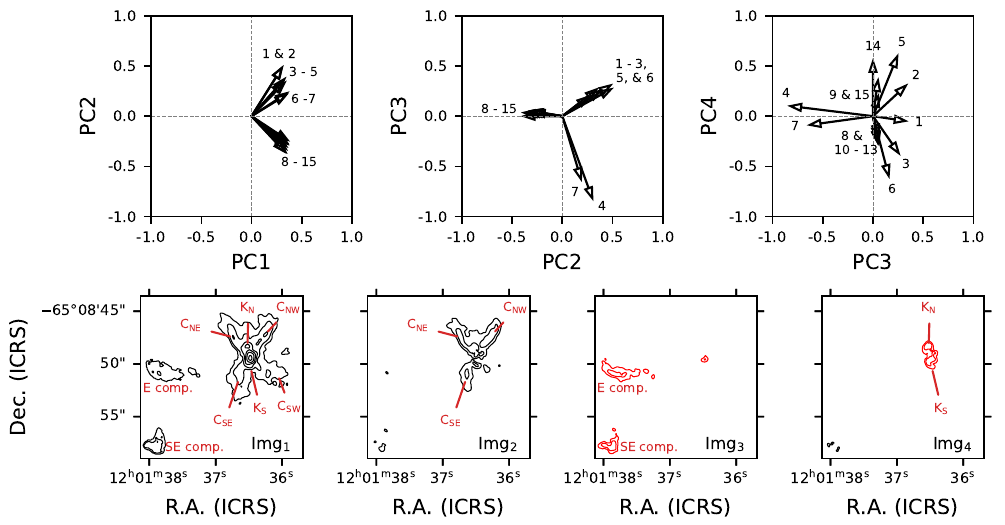}
\caption{Top: correlation wheel plots for the first four meaningful principal components of the PCA2D analysis. The labeled numbers for arrows correspond to the extended CH$_3$OH lines. Bottom: eigenimages (Img$_i$) for the meaningful PCs. The black and red contours represent the positive and negative contributions, respectively. Some representative positions, such as the outflow cavities (C$_\mathrm{NE}$, C$_\mathrm{NW}$, C$_\mathrm{SE}$, and C$_\mathrm{SW}$), northern and southern knots (K$_\mathrm{N}$ and K$_\mathrm{S}$), and two additional components E and SE, are annotated on the eigenimages.}
\label{fig:PCA2D_rst}
\end{figure*}

To interpret PC1, PC2, PC3, and PC4, we examined (i) the correlations among lines described by the component scores of each PC, and (ii) the spatial distribution of their contributions visualized as eigenimages ($\mathrm{Img}_i$). In Fig.~\ref{fig:PCA2D_rst}, the top panels show the correlation wheel plots visualizing the correlations among the lines (each arrow corresponds to the individual extended CH$_3$OH line, positioned by the component scores of two PCs), and the bottom panels show $\mathrm{Img}_i$ for PC1 to PC4, with black and red contours indicating positive and negative contributions, respectively. Each PC captures a distinct correlation pattern:
\begin{itemize}
\item PC1: All component scores are positive and of similar magnitude, indicating a strong positive correlation across the CH$_3$OH lines with extended emission. Accordingly, $\mathrm{Img}_1$ approximates a composite map of the emitting gas: the central component is clearly visible together with the northeast, northwest, and southeast cavity walls (C$_\mathrm{NE}$, C$_\mathrm{NW}$, C$_\mathrm{SE}$), and the northern and southern knots (K$_\mathrm{N}$, K$_\mathrm{S}$). The southwest cavity wall (C$_\mathrm{SW}$) is comparatively faint. Two additional emission features are seen to the east (E) and southeast (SE) of IRS1.

\item PC2: $\mathrm{Img}_2$ enhances the cavity walls while suppressing the contribution on the central region relative to $\mathrm{Img}_1$. Component scores span roughly $-0.4$ to $+0.4$, consistent with a component that encodes the relative prominence of the walls in the peak intensity maps: lines with positive component scores (lines 1–7) strengthen as their absolute values increase, whereas lines with negative component scores (lines 8–15) weaken with increasing absolute values.

\item PC3: $\mathrm{Img}_3$ shows no distinct structure near IRS1, but exhibits strong negative contributions toward the E and SE components. Lines 4 and 7 have significant negative component scores, indicating they are preferentially detected in E and SE. The nondetection of other CH$_3$OH lines at these positions suggests that the physical and/or excitation conditions in E and SE differ from those along the outflow cavity walls.

\item PC4: Only K$_\mathrm{N}$ and K$_\mathrm{S}$ are prominent in $\mathrm{Img}_4$, with negative contributions. Lines 5 and 14 have large positive component scores, whereas lines 3 and 6 have large negative component scores. Therefore, PC4 captures deviations in the relative line strengths between the knots and the cavity walls (the latter trend being primarily described by PC2; cf. K$_\mathrm{N}$ in Fig.~\ref{fig:mom8}). Lines 3 and 6 (large negative component scores) are brighter in the knots than other CH$_3$OH lines of comparable intensity along the walls (e.g., line 4), while lines 5 and 14 are comparatively weaker. Notably, lines 3 and 6 have higher upper state energies ($E_\mathrm{up}=37.8$ and $49.4$ K) than lines 5 and 14 ($E_\mathrm{up}=16.9$ and $27.0$ K), suggesting that PC4 traces elevated excitation temperatures in K$_\mathrm{N}$ and K$_\mathrm{S}$ relative to the cavity walls.
\end{itemize}

\subsection{Physical properties derived from methanol}
\subsubsection{Rotation diagram analysis} \label{sec:rot_diag}
\begin{figure*}
\centering
\includegraphics[width=\textwidth]{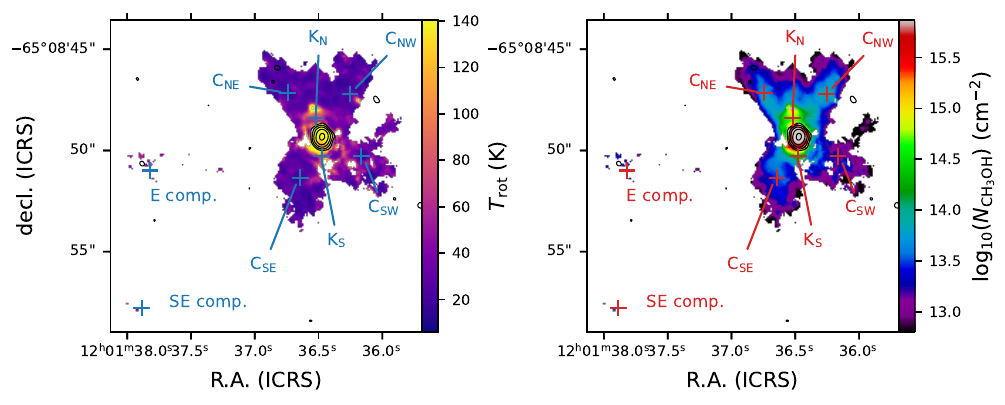}
\caption{Maps of the rotational temperature ($T_\mathrm{rot}$; left) and molecular column density ($N_\mathrm{CH_3OH}$; right) derived from the detected isolated CH$_3$OH lines (corresponding to the dashed orange lines in Fig.~\ref{fig:rot_rst2}). Black contours indicate the integrated intensity map of the PC1-weighted cube, representing the central component (the right panel of Fig.~\ref{fig:corino_mom0}) with the levels beginning from 5$\sigma$ and increasing by a factor of 2 (5$\sigma$, 10$\sigma$, 20$\sigma$, and so forth, where $\sigma = 0.21$~K~km~s$^{-1}$). The plus symbols on both maps mark the representative positions shown in Fig.~\ref{fig:PCA2D_rst}.
}
\label{fig:rot_rst}
\end{figure*}

\begin{figure*}
\centering
\includegraphics[width=\textwidth]{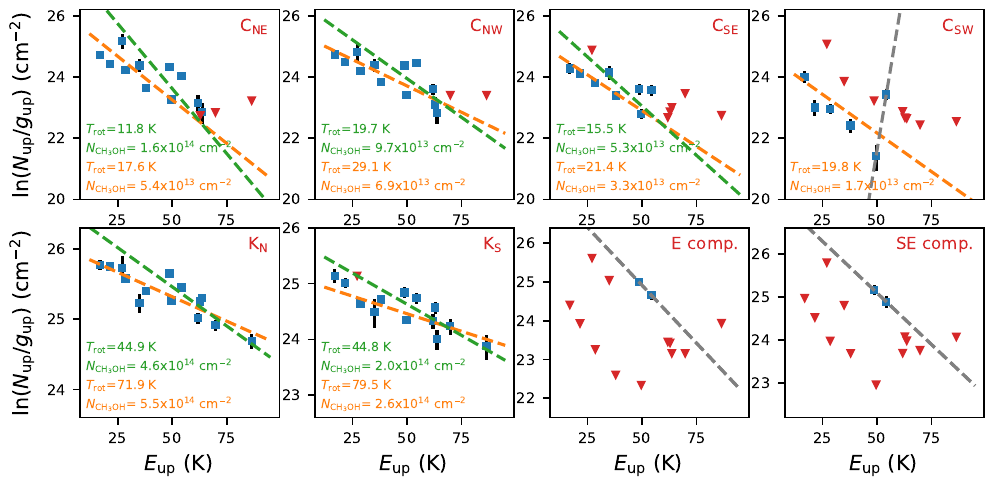}
\caption{Rotation diagrams of representative positions. Red triangles indicate the 3$\sigma$ upper limits of undetected lines. Fitting results using all isolated CH$_3$OH lines (dashed orange lines) and those restricted to transitions with $E_\mathrm{up}\ge40$~K (dashed green lines) are summarized in the bottom left corner of each panel. Fits considered unreliable are shown as dashed gray lines.
}
\label{fig:rot_rst2}
\end{figure*}

We performed a rotation-diagram analysis at positions where more than two isolated CH$_3$OH lines were detected. The line at 302.912979~GHz was excluded because it has been identified as a maser transition \citep{Jeo25compass}. Figure~\ref{fig:rot_rst} presents the derived $T_\mathrm{rot}$ and $N_\mathrm{CH_3OH}$ maps. Pixels flagged as inaccurate, or where the $1\sigma$ uncertainty in $T_\mathrm{rot}$ exceeds its measured value, are masked. The results toward the central component (outlined by black contours) are also unreliable, as the CH$_3$OH lines there are optically thick \citep{Aud25compass}. Figure~\ref{fig:rot_rst2} shows the rotation diagrams for the eight representative positions. For these, we additionally tested the effect of excluding low-excitation lines ($E_\mathrm{up}\le40$ K), which are prone to being optically thick.

Along the outflow cavity walls, $T_\mathrm{rot}$ ranges from 10 to 30 K and $N_\mathrm{CH_3OH}$ from $1\times10^{13}$ to $8\times10^{13}$ cm$^{-2}$. The results for C$_\mathrm{NE}$, C$_\mathrm{NW}$, and C$_\mathrm{SE}$ are consistent. Excluding the low-excitation lines yields slightly lower $T_\mathrm{rot}$s, but the change is minor compared to the uncertainties ($\sim$5 K). In contrast, the results for the southwest wall are highly uncertain due to very weak CH$_3$OH emission ($\lesssim1$~K peak intensity), which prevents reliable constraints on the physical conditions.

A striking feature in Fig.~\ref{fig:rot_rst} is seen toward K$_\mathrm{N}$ and K$_\mathrm{S}$. Here, $T_\mathrm{rot}$ reaches $\sim$70–80 K, with uncertainties of $\sim$5 K (K$_\mathrm{N}$) and $\sim$20 K (K$_\mathrm{S}$). The column densities range from $2\times10^{14}$ to $6\times10^{14}$ cm$^{-2}$, significantly higher than along the cavity walls, especially in K$_\mathrm{N}$. These elevated temperatures are consistent with the PC4, namely, stronger emission from higher-excitation lines (Sect.~\ref{sec:PCA2D}). When low-excitation lines are excluded, the fit in K$_\mathrm{N}$ yields $T_\mathrm{rot}=44.8\pm3.8$ K, clearly different from the full-line result. In this case, the low-excitation ($E_\mathrm{up}\le40$ K) transitions lie below the best-fit line, implying that some are optically thick. We need to verify the actual optical depths for these lines using spectral fitting to obtain accurate parameter determinations at K$_\mathrm{N}$ and K$_\mathrm{S}$. 

We also examine the E and SE components. For these, $T_\mathrm{rot}\approx20$ K and $N_\mathrm{CH_3OH}\approx3\times10^{14}$ cm$^{-2}$. However, only two transitions within a narrow $E_\mathrm{up}$ range are available, and the $3\sigma$ upper limits of nondetections fall well below the best-fit line. Thus, the derived parameters for E and SE remain highly uncertain.
 
\subsubsection{Spectral modeling results}
The spectral modeling of the CH$_3$OH lines showing extended emission was performed to evaluate the reliability of $T_\mathrm{rot}$ and $N_\mathrm{CH_3OH}$ derived from the rotation diagrams. We also included the overlapping CH$_3$OH lines in this spectral modeling. Extensive exploration with varying emission size revealed that numerous models can adequately reproduce the observed lines, suggesting considerable degeneracy for the cavity walls (C$_\mathrm{NE}$, C$_\mathrm{NW}$, C$_\mathrm{SE}$, and C$_\mathrm{SW}$) and knots (K$_\mathrm{N}$ and K$_\mathrm{S}$). We therefore assessed the best-fit models in two extreme cases: (1) fully resolved and (2) unresolved cases. The fully resolved model was derived by fixing the emission size to 10\arcsec. For the unresolved case, we selected the model with the smallest emission size among the local minima in the $\chi^2$ distribution. Table \ref{tbl:cassis} shows the best-fit parameters for the model spectra. 

\renewcommand{\arraystretch}{1.4}
\begin{table*}
\caption{CASSIS fitting results.}
\label{tbl:cassis}
\centering
\begin{tabular}{c c c c c c c}
\hline\hline
  Position & Emission size& FWHM& $T_\mathrm{ex}$ & $N_\mathrm{CH_3OH}$ & $\tau$\tablefootmark{a}& Reduced $\chi^2$\\
  & (\arcsec) & (km s$^{-1}$) & (K) & (cm$^{-2}$) & &\\
\hline
C$_\mathrm{NE}$&   10.00&	0.67&  27.69$^{+1.63}_{-1.55}$&	5.26$^{+0.24}_{-0.22}\times$10$^{13}$&    0.116& 0.711\\
C$_\mathrm{NE}$&	0.19&	0.58&  22.72$^{+1.49}_{-1.57}$&	5.55$^{+1.17}_{-1.07}\times$10$^{14}$&    2.192& 0.689\\
C$_\mathrm{NW}$&   10.00&	0.69&  32.46$^{+1.63}_{-1.74}$&	6.46$^{+0.26}_{-0.27}\times$10$^{13}$&    0.102& 0.766\\
C$_\mathrm{NW}$&	0.19&	0.61&  27.43$^{+2.09}_{-1.77}$&	6.08$^{+1.21}_{-1.00}\times$10$^{14}$&    1.732& 0.750\\
C$_\mathrm{SE}$&   10.00&	0.64&  27.16$^{+2.17}_{-2.10}$&	3.44$^{+0.21}_{-0.20}\times$10$^{13}$&    0.084& 0.670\\
C$_\mathrm{SE}$&	0.19&	0.59&  24.78$^{+2.61}_{-2.33}$&	3.06$^{+1.03}_{-1.20}\times$10$^{14}$&    1.091& 0.666\\
C$_\mathrm{SW}$&   10.00&	0.58&  31.10$^{+6.60}_{-4.66}$&	1.82$^{+0.22}_{-0.20}\times$10$^{13}$&    0.039& 0.600\\
C$_\mathrm{SW}$&	0.10&	0.44&  22.59$^{+3.82}_{-3.12}$&	9.53$^{+3.36}_{-2.47}\times$10$^{14}$&    5.156& 0.587\\
K$_\mathrm{N}$&	   10.00&	1.61&  99.62$^{+3.06}_{-2.80}$&	7.93$^{+0.26}_{-0.27}\times$10$^{14}$&    0.050& 0.837\\
K$_\mathrm{N}$&	    0.20&	1.58&  93.50$^{+3.16}_{-3.13}$&	5.29$^{+0.74}_{-0.60}\times$10$^{15}$&    0.399& 0.831\\
K$_\mathrm{S}$&	   10.00&	1.39& 117.87$^{+6.74}_{-6.84}$&	4.67$^{+0.32}_{-0.33}\times$10$^{14}$&    0.023& 0.730\\
K$_\mathrm{S}$&	    0.13&	1.35& 108.52$^{+6.35}_{-6.08}$&	7.03$^{+2.33}_{-1.59}\times$10$^{15}$&    0.481& 0.725\\
\hline
\end{tabular}
\tablefoot{
\tablefoottext{a}{The optical depth for the thickest line in the best-fit model.}
}
\end{table*}
\renewcommand{\arraystretch}{1.0}

\begin{figure*}
\centering
\includegraphics[width=0.85\textwidth]{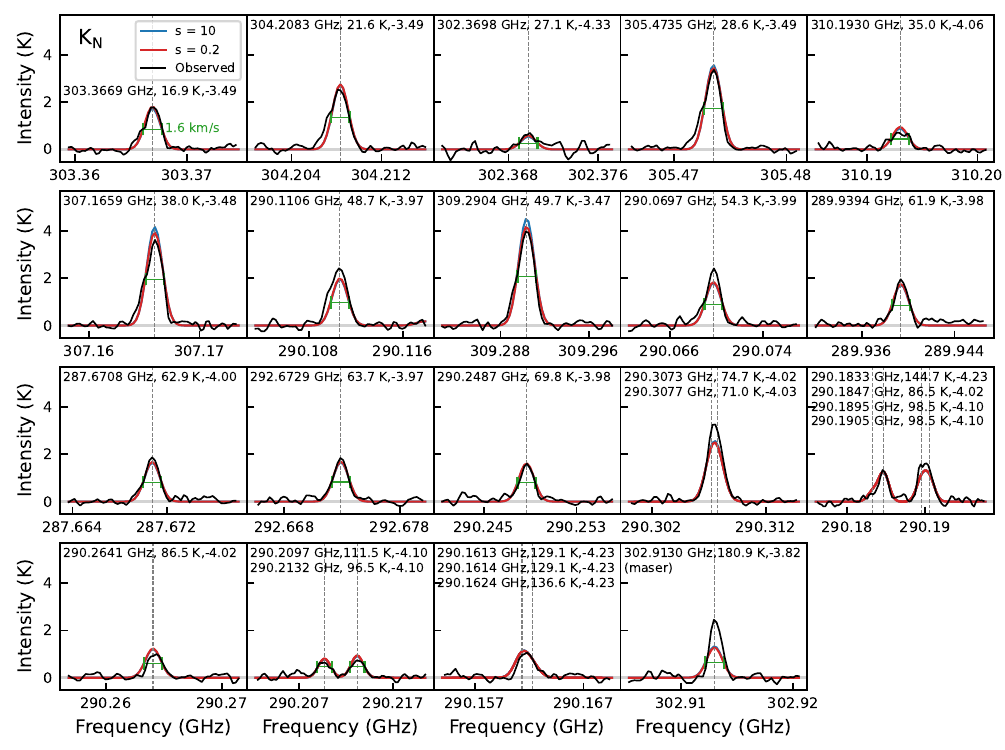}
\caption{Best-fit model spectra for the CH$_3$OH lines detected at K$_\mathrm{N}$. The fully resolved model ($s = 10\arcsec$) and the unresolved model ($s =$ 0\farcs2) are shown in blue and red, respectively. The vertical dashed gray lines indicate the rest frequencies of the CH$_3$OH lines. In panels without significant line blending, horizontal green lines indicate the full width at half-maximum of the fully resolved model spectra. In most cases, the red and blue spectra overlap, so the red spectra are mostly visible in the panels. The rest frequencies, $E_\mathrm{up}$ (K), and $\log_{10}(A_{\rm ij}\ (\mathrm{s}^{-1}))$ values for the CH$_3$OH lines are indicated in each panel.}
\label{fig:CASSIS}
\end{figure*}

Figure~\ref{fig:CASSIS} shows the model spectra in both cases for K$_\mathrm{N}$ as an example. The model spectra for the other positions are available on Zenodo (Figs. C.1 to C.5). The model spectra for the fully resolved and unresolved cases are almost indistinguishable at all positions. In C$_\mathrm{NE}$, C$_\mathrm{NW}$, C$_\mathrm{SE}$, and C$_\mathrm{SW}$, only a small decrease in peak intensities is found for a few transition lines, where these lines become optically thick ($\tau >$ 1) in the unresolved models. However, this difference is still negligible; the reduced $\chi^2$ values for the two models differ only by a few percent (see Table \ref{tbl:cassis}). The indistinguishable models in these two extreme cases suggest numerous possible solutions spanning a wide range of emission sizes. 

The spectral modeling for the E and SE components fails to produce an acceptable model. None of the LTE model spectra can simultaneously account for the detected CH$_3$OH lines and the absence of others. This implies that the non-LTE effect likely applies in these components. We subsequently discuss this later in Sect.~\ref{sec:non_LTE}.

\section{Discussion} \label{sec:discuss}
\begin{figure*}
\centering
\includegraphics[width=1.0\textwidth]{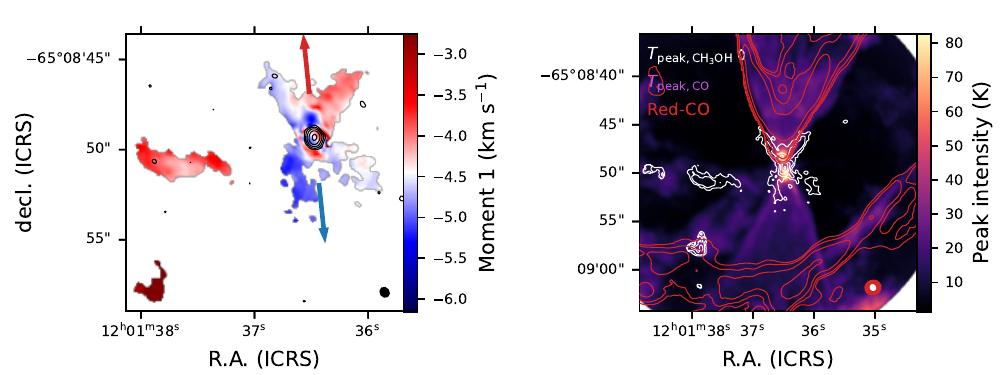}
\caption{Left: intensity-weighted velocity (moment 1) map of the CH$_3$OH line at 290.110637~GHz (line 4). The black ellipse on the bottom right corner represents the beam size (0\farcs52 $\times$ 0\farcs44). Black contours show the integrated PC1 map representing the central component (the bottom panel of Fig.~\ref{fig:corino_mom0}). The contour levels begin from 5$\sigma$ and increases by a factor of two (5$\sigma$, 10$\sigma$, 20$\sigma$, and so forth, where $\sigma = 0.21$~K~km~s$^{-1}$). The red and blue arrows indicate the directions of the large-scale CO outflows. The color scale of the moment 1 map is limited to $\pm1.7$~km~s$^{-1}$ from $v_\mathrm{LSR}$ to highlight the velocity gradient along the outflow cavity walls. Right: comparison between CH$_3$OH and $^{12}$CO emission. The white contours represent the peak-intensity map for line 4 ($T_\mathrm{peak,CH_3OH}$), while the background image displays the peak intensity map of CO ($J=2$--1) from the ALMA eDisk Large Program \citep[$T_\mathrm{peak,CO}$][]{Gav24}. The integrated intensity map of red-shifted $^{12}$CO outflows \citep[Red-CO;][]{Zap18}, integrated from 0 to +20~km~s$^{-1}$, is shown as red contours. The red and white ellipses in the bottom right corner represent the beam sizes of the Red-CO and $T_\mathrm{peak,CH_3OH}$ maps, respectively. Note that $T_\mathrm{peak,CO}$ map has been convolved to the CH$_3$OH beam size. The contour levels are 5$\sigma$, 10$\sigma$, 20$\sigma$, and so forth, where $\sigma = 0.12~K$ and 0.55~K~km~s$^{-1}$ for the $T_\mathrm{peak,CH_3OH}$ and Red-CO maps, respectively.}
\label{fig:edisk}
\end{figure*}

The extended emission structures can be divided into three physically distinct components: the outflow cavity walls, the knots, and the E-SE features. Each region exhibits characteristic morphologies and excitation conditions, as revealed by the PCA analysis and supported by both the rotation diagram and spectral modeling results. In the following subsections, we discuss the physical interpretation of each component in detail, emphasizing how the CH$_3$OH emission traces different dynamical and chemical processes. We also assess plausible methanol desorption mechanisms based on the derived physical parameters. The kinematics of the extended structures were evaluated from the moment 1 map of line~4, which is detected across all components (left panel of Fig.~\ref{fig:edisk}). Finally, we compared the morphology of the extended CH$_3$OH emission with archival ALMA $^{12}$CO $J=$2$-$1 data: one revealing the overall structure of bipolar outflows from IRS1 and IRS2 at an angular resolution of $\sim$ 1\farcs5 \citep{Zap18}, and another providing a detailed view of the outflows near IRS1 with higher angular resolution \citep[$\sim$ 0\farcs09; from the eDisk Large Program; ][]{Gav24}.

\subsection{Outflow cavity walls} \label{sec:cavity}
The outflow cavity walls are clearly distinguished as extended CH$_3$OH emitting structures with narrow line widths of $\sim$0.7~km~s$^{-1}$. PC2 of the PCA2D analysis clearly reveals their spatial extent. The elongated wall structures in the northern cavity are distinct and extend out to $\sim$3\farcs5, corresponding to a deprojected distance of about 800~au from the continuum peak, assuming an inclination angle of 130\degr\ \citep{Yan20} and a distance of 176~pc \citep{Jes25compass}. Along the cavity walls, several methanol lines (particularly lines 1--7) with low $E_\mathrm{up}$ and high $A_\mathrm{ij}$ trace these elongated structures. The moment 1 map reveals that these X-shaped cavity walls have a shallow velocity gradient perpendicular to the outflow direction, indicating rotation. In addition, the CH$_3$OH cavity walls enclose the molecular outflows traced by $^{12}$CO $J=2$--1. Both results imply that the CH$_3$OH lines are emitted from outflow cavity walls in the rotating envelope of IRS1.

\begin{figure*}
\centering
\includegraphics[width=1.0\textwidth]{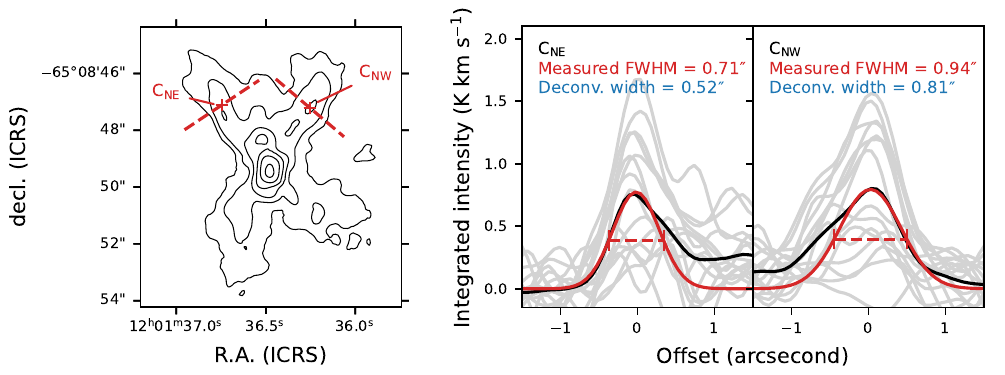}
\caption{Left: position of the spatial cuts (dashed red lines) centered on the C$_\mathrm{NE}$ and C$_\mathrm{NW}$ positions. The black contours show the map of $Img_1$. Right: integrated intensity profiles extracted from the two spatial cuts in the left panel. The intensity profiles are derived from the integrated intensity maps derived with the moment-masking technique. The gray lines represent the integrated intensity profiles of individual CH$_3$OH lines, and the black line represents their average profile. The red line shows the Gaussian-fitting results for the cavity walls, excluding the plateaus between these cavities. The measured and deconvolved FWHM sizes are indicated in red and blue, respectively.}
\label{fig:C_width}
\end{figure*}

Neither the rotation diagram nor the CASSIS spectral-fitting analysis alone can tightly constrain $N_\mathrm{CH_3OH}$, because the emission size is unknown. We therefore estimated the transverse widths of the observed cavity structures directly from the images. The northern cavity walls are more clearly detected than their southern counterparts, allowing reliable width measurements toward C$_\mathrm{NE}$ and C$_\mathrm{NW}$. Figure~\ref{fig:C_width} shows the transverse profiles of integrated intensity. The deconvolved widths measured from these profiles are 0\farcs52 and 0\farcs81 for C$_\mathrm{NE}$ and C$_\mathrm{NW}$, respectively. Both widths rule out the case of unresolved emission.

Using the deconvolved widths as emission sizes in the CASSIS modeling, we obtain best-fit excitation temperatures of $T_\mathrm{ex}=27.2$~K for C$_\mathrm{NE}$ and $32.3$~K for C$_\mathrm{NW}$, consistent with both the fully resolved case and the rotational temperatures ($T_\mathrm{rot}$) from the rotation diagrams. The best-fit models for C$_\mathrm{NE}$ and C$_\mathrm{NW}$ also indicate that all detected lines are optically thin, with $\tau < 0.23$ and $0.14$, respectively.

To estimate the true CH$_3$OH column density, we accounted for the filling factor, defined as the fraction of the Gaussian beam area occupied by the Gaussian-shaped line-emitting region:
\begin{equation}
\eta \;=\; \frac{\theta_\mathrm{source}^2}{\theta_\mathrm{source}^2 + \theta_\mathrm{beam}^2},
\end{equation}
where $\theta_\mathrm{source}$ and $\theta_\mathrm{beam}$ are the FWHM sizes of the line-emitting region and the beam, respectively. In the fully resolved case ($\theta_\mathrm{source}=10\arcsec$), $\eta\approx1$. With the deconvolved widths, the values are $\eta=0.538$ and $0.738$ for C$_\mathrm{NE}$ and C$_\mathrm{NW}$, respectively. Assuming the line-emitting gas has bar-like geometry with the measured widths, the actual filling factors are expected to lie between these $\eta$ values and unity. Consequently, the true CH$_3$OH column densities fall between those inferred for the fully resolved case and those derived using the computed $\eta$. For C$_\mathrm{NE}$, $N_\mathrm{CH_3OH}$ ranges from $5.3\times10^{13}$ to $9.9\times10^{13}$~cm$^{-2}$, and for C$_\mathrm{NW}$, from $6.5\times10^{13}$ to $8.8\times10^{13}$~cm$^{-2}$. In summary, the methanol column density in the cavity walls is estimated to be $(5$–$10)\times10^{13}$~cm$^{-2}$. Note that the derived column density is much lower than that of the hot corino of IRS1, estimated by performing the spectral-modeling analysis in the continuum peak position \citep[$\sim 5 \times 10^{19}$~cm$^{-2}$; ][]{Aud25compass}. 

\subsection{Northern and southern knots} \label{sec:knots}
The northern and southern knots, identified in PC1 and PC4 of the PCA2D results, appear as compact structures located north and south of the central component. The northern knot is particularly prominent: all 26 extended CH$_3$OH lines are detected with line widths of $\sim$1.5~km~s$^{-1}$ (Fig.~\ref{fig:CASSIS}), and it shows a well-defined compact morphology in the peak intensity maps (Fig.~\ref{fig:mom8}). This feature lies $\sim$1\arcsec\ north of the continuum peak, corresponding to a deprojected distance of $\sim$180~au. In contrast, the southern knot is located very close to the central component, making it difficult to separate the two.

The kinematics of the knots differ from those of the cavity walls. In the moment 1 map (left panel of Fig.~\ref{fig:edisk}), both knots exhibit line-of-sight velocities opposite to those of the large-scale CO outflows. Relative to the central component, the northern knot is blue-shifted ($-0.85$~km~s$^{-1}$) and the southern knot is red-shifted ($+0.75$~km~s$^{-1}$). Similar opposite-velocity features at comparable positions were seen in HCN $J=4$--3 by \citet{Yan20}: the northern CH$_3$OH knot coincides spatially with their N2 knot, whereas the southern CH$_3$OH knot lies slightly farther south than their S1 knot. We note, however, that the CH$_3$OH knot velocities differ from those of the HCN knots, likely owing to strong self-absorption of HCN near the systemic velocity. In addition, the reported size of the HCN N2 knot ($0\farcs058$; \citealt{Yan20}) is much smaller than the emission size derived from the unresolved CH$_3$OH model ($0\farcs2$). Therefore, we cannot determine whether the N2 knot and the northern CH$_3$OH knot share a common origin. \citet{Yan20} proposed two possible origins for such knots: (1) parts of the outflow and (2) infalling gas behind the cavity walls.

Spectral fits with the unresolved model indicate that all CH$_3$OH lines remain optically thin ($\tau<1$), implying that low-excitation transitions ($E_\mathrm{up}\le40$~K) can be included in the rotation-diagram analysis. In principle, this should yield results consistent with those from fully resolved spectral fitting. In practice, the rotation diagrams give $T_\mathrm{rot}\sim70$--80~K, whereas spectral fitting yields $T_\mathrm{ex}\sim100$--120~K. Overlaying the spectral-fitting solution on the rotation diagram shows that both reproduce the observed data reasonably well; therefore, we adopte a temperature range of $T_\mathrm{ex}=70$--120~K. The derived $N_\mathrm{CH_3OH}$ spans $10^{14}$--$10^{16}$~cm$^{-2}$, depending on the assumed emission size. Both $T_\mathrm{ex}$ and $N_\mathrm{CH_3OH}$ exceed those of the cavity walls, suggesting that the knots experience different sublimation and/or excitation conditions than the walls.

\subsection{East and southeast components}
Both the E and SE components are highlighted in PC3, indicating that only two CH$_3$OH lines at 290.110637 and 290.069747~GHz trace these features. Rotation-diagram analysis yields $T_\mathrm{rot}\sim15$–20~K and $N_\mathrm{CH_3OH}\sim3\times10^{14}$~cm$^{-2}$, but these estimates are highly uncertain. Additionally, LTE spectral modeling does not accurately reproduce the observed spectra.

The E and SE components exhibit red-shifted velocities relative to IRS1 ($1.0$ and $2.4$~km~s$^{-1}$, respectively) and lie farther from IRS1 than the extent of its CO outflows (right panel of Fig.~\ref{fig:edisk}). From the archival $^{12}$CO data, we find no corresponding CO emission toward E component, leaving its relationship to IRS1 unclear. On the other hand, the CO counterpart to the SE component is clearly identified. Additionally, the SE component is located at the boundary of the red-shifted outflow from IRS2, which is deflected by the blue-shifted outflow of IRS1 \citep{Zap18}. These kinematic and morphological features imply that the SE component is associated with the outflow of IRS2 rather than IRS1. Consequently, we assume that the E and SE components are not directly related to IRS1 and exclude them from further detailed discussion.

\subsection{Plausible desorption mechanisms}
\citet{Dro15} modeled envelopes and cavity walls of low-mass protostars with a stellar luminosity of 35.7~$L_\odot$. Their results showed that reactive desorption can actively release many COMs in cavity-wall regions, even with an efficiency of only 1\%. FUV photons from the central protostar can penetrate the cavity walls, photodissociating CH$_3$OH on grain surfaces; subsequent recombination leads to frequent reactive desorption, injecting CH$_3$OH into the gas phase. In the inner cavity wall ($z<1000$~au) in this model, the predicted CH$_3$OH abundance exceeds $10^{-8}$. Gas-phase CH$_3$OH then persists for $\sim10^3$~yr before photodissociating. This timescale is likely a lower limit, because this specific physical model is static; continuous replenishment of fresh dust and ice through envelope infall would prolong the gas-phase lifetime of CH$_3$OH.

\citet{Bur19} investigated dust sputtering in cold cores subject to C-type shocks. For pre-shock densities of $n_\mathrm{H_2}\sim5\times10^4$~cm$^{-3}$ and shock velocities of 20~km~s$^{-1}$, their models showed that CH$_3$OH can be efficiently ejected from grain mantles by sputtering. The resulting gas-phase abundance reaches $\sim10^{-7}$ and is maintained for $\sim10^4$~yr until freeze-out dominates.

\citet{Jes25compass} estimate the spatial extent of warm ($T>100$~K) gas heated by the central source in an envelope with a power-law density profile $\rho\propto r^{-3/2}$; the expected radius for BHR71-IRS1 is $\sim$52~au. This size is roughly 3 times smaller than the observed extent, probably due to an unresolved high density structure at the center \citep{Kim25}. Comprehensive consideration, including the models of \citet{Dro15} and \citet{Bur19}, suggests that thermal desorption is unlikely to operate efficiently across most of the extended structures in BHR71-IRS1. Although radiative heating can propagate along the outflow cavities, the 35.7~$L_\odot$ model of \citet{Dro15} still yields $T_\mathrm{dust}\ll100$~K in the cavity walls. C-type shock models likewise predict little increase in $T_\mathrm{dust}$ \citep{Bur19}. Taken together, these considerations support the need for nonthermal desorption processes to explain the extended emission.

Among the nonthermal mechanisms, reactive desorption and dust sputtering in C-type shocks are the most plausible to account for the observed CH$_3$OH. In the following sections, we assess their applicability by comparing model predictions with the observed morphologies, kinematics, and excitation conditions in the cavity walls and knot structures of BHR71-IRS1.

\subsubsection{Reactive desorption}
The bolometric luminosity of BHR71-IRS1 ($9.2\,L_\odot$) is lower than the stellar luminosity adopted by \citet{Dro15}. They explored luminosity dependence and showed that a 15~$L_\odot$ model yields dust temperatures $T_\mathrm{dust}$ cooler by at most $\sim$5~K, with CH$_3$OH abundances broadly consistent with the fiducial (35.7~$L_\odot$) case, but with a spatial shift. For comparison, we assumed that the envelope of BHR71-IRS1 has chemical abundances comparable to those in the reactive-desorption model with the luminosity of 35.7~$L_\odot$.

Within the inner cavity wall ($z<1000$~au), $T_\mathrm{dust}$ is 15–35~K when accounting for the $\sim$5~K offset between the 35.7 and 15~$L_\odot$ models. This range is consistent with the CH$_3$OH excitation temperatures inferred along the cavity walls. The reactive-desorption models predict typical CH$_3$OH abundances ($n_\mathrm{CH_3OH}/n_\mathrm{H}$) of $10^{-8}$–$10^{-7}$ along the walls. Because our data do not directly constrain the absolute CH$_3$OH abundance, we instead scaled the best-fit $N_\mathrm{CH_3OH}$ by the relative abundances of other COMs \citep[from the chemical model of ][]{Dro15} to construct synthetic spectra using CASSIS for comparison with the observed spectra toward C$_\mathrm{NE}$.

Unless otherwise noted, CASSIS parameters followed the fully resolved best-fit solution for C$_\mathrm{NE}$ (Table~\ref{tbl:cassis}). Molecular column densities of other COMs were obtained by multiplying the best-fit $N_\mathrm{CH_3OH}$ by their relative abundances. In most tests, the synthetic spectra do not show lines above the noise level, consistent with the observations. The only notable exception is CH$_3$CHO: with a predicted abundance of $10^{-8}$–$10^{-7}$, the conservative consideration (assuming a CH$_3$OH abundance of 10$^{-8}$) with a relative abundance of 10 produces detectable lines, whereas the unity case does not. Given the plausible ranges for CH$_3$OH and CH$_3$CHO, these spectral tests do not significantly contradict the observations; reactive desorption remains a viable mechanism for the cavity walls.

In contrast, reactive desorption is unlikely to account for the gas-phase CH$_3$OH observed in the knot structures. \citet{Yan20} outlined the line-of-sight geometry of the cavity walls: in the north, the near-side wall lies close to the plane of the sky, whereas the far-side wall starts along the line of sight and gradually shifts northward. Adopting this geometry, the position of the HCN N2 knot (coincident with the northern CH$_3$OH knot) corresponds to either (1) the near-side wall at an offset of $\sim$100~au from the protostar or (2) the far-side wall at $\sim$400~au. Infall on the near-side wall is inconsistent with the clearly blue-shifted velocity of the northern knot relative to IRS1. A blue shift is feasible on the far-side wall, but the measured CH$_3$OH temperatures ($T_\mathrm{ex}\approx70$–120~K) cannot be achieved by radiative heating from the central source at those locations \citep{Dro15,Jes25compass}. We therefore reject reactive desorption as a viable explanation for the CH$_3$OH emission in the knots.

\subsubsection{Dust sputtering by a C-type shock}

\citet{Bur19} showed that dust and gas velocities become decoupled in C-type shocks, with their relative velocity increasing over the first $\sim$100~yr. CH$_3$OH ice is rapidly sputtered once the relative velocity becomes sufficient, leading to a sudden rise in the gas-phase CH$_3$OH abundance up to the pre-sputtering ice levels. The gas temperature ($T_\mathrm{gas}$) begins to rise after $\sim$100~yr. If the observed excitation temperature is assumed to represent $T_\mathrm{gas}$, cold gas-phase CH$_3$OH could exist in the very early stages of the shocked gas. However, $T_\mathrm{gas}$ increases rapidly, exceeding 200~K within $\sim$200~yr, which places a strong limitation on this early-stage scenario. The models also show that $T_\mathrm{gas}$ cools again after $\sim3\times10^3$~yr, allowing CH$_3$OH to be present in the cold gas phase. In this late stage, gas-phase CH$_3$OH can persist until freeze-out dominates ($\sim10^4$~yr). Because the late-stage scenario provides a longer timescale for survival, we assessed its applicability to the extended structures.

Applying the C-type shock scenario to the cavity walls is less straightforward. Although previous studies suggested that past low-velocity ($\sim$20~km~s$^{-1}$) outflow events triggered partially dissociative shocks in distant gas \citep{Gus15,Ben17} approximately 4000~years ago, and that these outflows interacted with the cavity walls, several arguments disfavor this interpretation. The narrow line widths observed in the cavity walls ($\sim$0.7~km~s$^{-1}$) suggest a relatively quiescent kinematic environment. Moreover, the spatial discrepancy between the CO outflows and cavity walls further weakens the shock scenario. Episodic shocks caused by bipolar outflows may also be inconsistent with the continuous CO outflow structures (see Fig.~\ref{fig:edisk}). Therefore, we rule out C-type shocks as the primary mechanism for CH$_3$OH emission in the cavity walls.

In contrast, the knot structures provide a more plausible site for shocks. The CH$_3$OH lines in the knots exhibit higher excitation temperatures (70--120~K) and broader line widths ($\sim$1.5~km~s$^{-1}$) compared to the cavity walls. Within the C-type shock framework, the observed emission can be explained as the remnant of a low-velocity shock event ($\sim3\times10^3$--$10^4$~yr ago). Furthermore, \citet{Jeo25compass} report a methanol maser candidate (the 302.9130~GHz line) in the knots: a collisionally pumped (Class I) maser that likely originates from a shock, providing further support for the shock scenario. 

\citet{Bur19} also modeled the chemical evolution of other COMs (CH$_3$CHO, CH$_3$COCH$_3$, CH$_3$OCH$_3$, HCOOCH$_3$, and NH$_2$CHO). We conducted the same spectral-modeling tests as for the cavity walls, adopting parameters from the fully resolved best-fit model of K$_\mathrm{N}$. Again, molecular column densities of these species were derived by multiplying the best-fit $N_\mathrm{CH_3OH}$ by their relative abundances with respect to CH$_3$OH. Except for NH$_2$CHO, the models predict no detectable lines, consistent with the observed line-free spectra. For NH$_2$CHO, the models produce lines stronger than observed, possibly due to uncertainties in its solid-phase abundance \citep{Bur19}. It is also possible that NH$_2$CHO is intrinsically deficient in BHR71-IRS1, since even the hot corino spectra show no strong NH$_2$CHO features. Except for this discrepancy, the shock-based models do not significantly contradict the data; therefore, we do not rule out the C-type shock scenario. Further studies, including detailed line identification and spectral modeling toward both the hot corino and the knots, will be necessary to constrain the dominant desorption mechanism.

For the distant shocked gas, \citet{Ben17} derived shock properties similar to those adopted by the model of \citet{Bur19}. However, regarding the origin of the shocks, large-scale outflows are unlikely, as line-of-sight velocities of the knots are opposite to those of the large-scale outflows. Instead, we propose that the shocks arise from infalling material impacting the envelope, which naturally explains the compact morphology and distinct kinematic signatures of the knots.

\subsection{Nonthermal excitation of CH$_3$OH} \label{sec:non_LTE}
Among the CH$_3$OH lines, transitions with low $E_\mathrm{up}$ and high $A_{\rm ij}$ are preferentially detected over more extended regions (Table~\ref{tbl:extended}). Based on this observational trend, line~1 would be expected to be the strongest in cold to warm gas ($T_\mathrm{ex}\lesssim100$~K) within the extended structures, followed by lines~2, 3, and so on. However, only two CH$_3$OH lines at 290.069747 and 290.110637~GHz (designated as lines~7 and~4, respectively) are detected toward the E and SE components. Their $E_\mathrm{up}$ are 48.7 and 54.3~K, neither of which correspond to the lowest-excitation lines.

Because no other CH$_3$OH lines are seen, the observed spectra cannot be reconciled with neither the rotation-diagram nor the LTE spectral-modeling analyses. Both approaches imply that, under LTE, additional CH$_3$OH lines should accompany lines~7 and~4; this is evident in the rotation diagrams (bottom right panels of Fig.~\ref{fig:rot_rst2}), where no single straight line can simultaneously satisfy the two detections and the $3\sigma$ upper limits on the nondetections. These results argue against the LTE assumption. To determine the physical conditions in the E and SE components, further analysis that includes non-LTE excitation is required.

\section{Summary} \label{sec:summary}
In this study, we identified CH$_3$OH emission that extends beyond the central hot corino of BHR71-IRS1 using ALMA Band~7 spectra from the COMPASS Large Program. Multiple extended CH$_3$OH transitions were identified and analyzed with rotation-diagram and LTE spectral-modeling methods to derive excitation temperatures and column densities. We then evaluated plausible desorption mechanisms for the gas-phase CH$_3$OH by comparing the observations with model predictions. The main results are as follows.

\begin{itemize}
    \item The central component of IRS1 is identified from five isolated, strong COM lines assumed to trace the same structure. The central component shows rotational motion, consistent with previous studies.
    \item CH$_3$OH emission is detected from both the central hot corino and extended structures near IRS1. Across the extended regions, a total of 26 CH$_3$OH lines are detected. Lines tend to be stronger and more extended when $E_\mathrm{up}$ is lower and $A_{\rm ij}$ is higher.
    \item The spatial distribution of CH$_3$OH emission is characterized using two-dimensional principal component analysis. Analysis of 15 isolated CH$_3$OH lines separates the extended emission into three distinct regions with different excitation conditions: (1) X-shaped cavity walls, (2) northern and southern knots, and (3) additional components located east and southeast of IRS1.
    \item The rotating cavity walls extend out to $\sim$800~au toward the northeast, northwest, southeast, and southwest. The northern knot, with a well-isolated compact structure, is located $\sim$180~au from IRS1. Across these extended structures, thermal desorption is unlikely to account for the CH$_3$OH emission, as the region containing hot ($T > 100$~K) gas is estimated to be confined to only the inner $\sim$~52~au.
    \item The cavity walls are traced by cold gas-phase CH$_3$OH emission with excitation temperatures of 20--30~K. Accounting for the deconvolved widths of the northern walls, the column density is estimated to be $(5$--$10)\times10^{13}$~cm$^{-2}$. Reactive desorption driven by FUV radiation from the central protostar is the most plausible mechanism for releasing CH$_3$OH from the ice into the gas phase. In contrast, dust sputtering is disfavored due to the narrow observed line widths ($\sim$0.7~km~s$^{-1}$).
    \item The northern and southern knots exhibit higher excitation temperatures (70--120~K) and column densities ($10^{14}$--$10^{16}$~cm$^{-2}$) than the cavity walls. They also exhibit distinct kinematics, with velocities opposite to those of the large-scale CO outflows, suggesting that they are not part of the cavity walls. The most plausible release mechanism is dust sputtering triggered by a C-type shock $\sim10^{4}$~yr ago, possibly induced by infalling material (e.g., a streamer) impacting the envelope. Future high-angular-resolution observations will be necessary to constrain emission sizes better and thereby refine temperature and column density estimates.
    \item For the eastern and southeastern components, where only two CH$_3$OH lines are detected, the physical parameters cannot be reliably constrained. The absence of low-excitation transitions prevents a self-consistent LTE solution. Future studies that include non-LTE excitation will be required to assess their origin in detail.
\end{itemize}

\section*{Data availability}
The additional figures are available in the Zenodo repository: \url{https://doi.org/10.5281/zenodo.19499024}.

\begin{acknowledgements}
This paper makes use of the following ALMA data: ADS/JAO.ALMA\#2022.1.00316.L. ALMA is a partnership of ESO (representing its member states), NSF (USA) and NINS (Japan), together with NRC (Canada), NSTC and ASIAA (Taiwan), and KASI (Republic of Korea), in cooperation with the Republic of Chile. The Joint ALMA Observatory is operated by ESO, AUI/NRAO and NAOJ. The project leading to this publication has received support from ORP, that is funded by the European Union’s Horizon 2020 research and innovation programme under grant agreement No 101004719 [ORP]. J.-E.L., J.-H.J., H.-S.Y., and C.-H.K. were supported by the National Research Foundation of Korea (NRF) grant funded by the Korea government (MSIT) (grant numbers RS-2024-00416859 and RS-2026-25490557). + J.-H.J was also supported by Basic Science Research Program through the National Research Foundation of Korea (NRF) funded by the Ministry of Education (grant number RS-2025-25402622). A.C. and F.C.-S.M. received financial support from the European Research Council (ERC) under the European Union’s Horizon 2020 research and innovation programme (ERC Starting Grant “Chemtrip”, grant agreement No 949278). Y.-L.Y. acknowledges support from a Grant-in-Aid from the Ministry of Education, Culture, Sports, Science, and Technology of Japan (20H05845, 20H05844), and a pioneering project in RIKEN (Evolution of Matter in the Universe). P.N. acknowledges support from the ESO and IAU Gruber Foundation Fellowship programs. S.M. acknowledges support from the Thematic Action “Physique et Chimie du Milieu Interstellaire” (PCMI) of INSU Programme National “Astro”, with contributions from CNRS Physique \& CNRS Chimie, CEA, and CNES. D.H. was supported by a Center for Informatics and Computation in Astronomy (CICA) grant and grant number 110J0353I9 from the Ministry of Education of Taiwan. + D.H. also acknowledges support from the National Science and Technology Council, Taiwan (Grant NSTC111-2112-M-007-014-MY3, NSTC113-2639-M-A49-002-ASP, and NSTC113-2112-M-007-027). A.K. was supported by the NKFIH NKKP grant ADVANCED 149943 and the NKFIH excellence grant TKP2021-NKTA-64 (project no.149943 has been implemented with the support provided by the Ministry of Culture and Innovation of Hungary from the National Research, Development and Innovation Fund, financed under the NKKP ADVANCED funding scheme) + A.K. was also supported by COST (European Cooperation in Science and Technology) through COST Action PLANETS CA22133. S.S. acknowledges the support of the Max Planck Society. S.Z. thanks RIKEN Special Postdoctoral Researcher Program (Fellowships) for financial support.
\end{acknowledgements}

\bibliographystyle{aa}
\bibliography{references}

\begin{appendix}
\section{Selecting meaningful principal components in PCA} \label{App:PCA2D}
We utilized PCA to assess the spatial distribution of the observed line intensities \citep{Ung97,Oko21,Yun23}. PCA decomposes the variation of the original data into multiple orthogonal principal components (PCs) by deriving the eigenvectors and eigenvalues of a correlation matrix. The eigenvectors ($\vec{u}_i$) indicate correlations among the spectral lines, while the eigenvalues ($\lambda_i$) quantify the fraction of the total variation represented by each PC. Low-order PCs with large $\lambda_i$ capture the prominent emission feature, while high-order PCs with small $\lambda_i$ mainly represent noise-dominated signals.

The variation in the spatial distributions of line intensities in the original data can be explained by a set of PCs that account for a significant fraction of the variance (i.e., meaningful PCs). When the correlation matrices are derived from three-dimensional spectral cubes, the resulting PCs capture the dominant variations in position-position-velocity space (hereafter, PCA3D). In contrast, when matrices are computed from two-dimensional images (e.g., integrated or peak intensity maps), the PCs highlight the dominant spatial structures (hereafter, PCA2D).

In Sect.~\ref{sec:corino}, we applied PCA3D to investigate the representative kinematic structure of the central component near IRS1. Five isolated COM lines listed in Table \ref{tbl:isol_str} were used, resulting in five PCs. Among them, PC1 has an eigenvalue $\lambda_1 = 3.901$, accounting for $\sim$ 80~\% of the total variance. This dominance is expected because the selected COM lines share a confined spatial distribution within the central component, implying a common origin. The corresponding eigenimage (PC1-weighted cube) thus effectively represents the structure of the central component in the position-position-velocity space.

\begin{figure}
\centering
\includegraphics[width=0.45\textwidth]{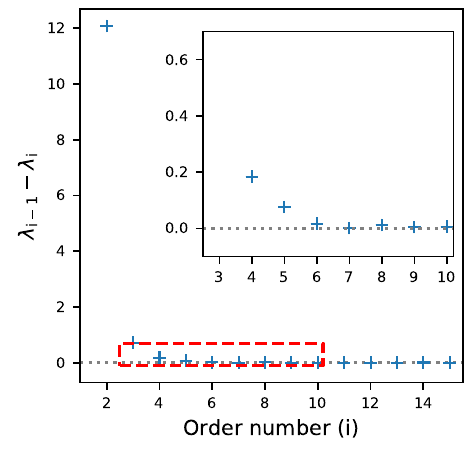}
\caption{Change of $\lambda_i$ ($\lambda_{i-1} - \lambda_i$) presented as a function of order number ($i$). An inset exhibits a zoomed-in view of the scree plot marked with the dashed red box.}
\label{fig:scree}
\end{figure}

In Sect.~\ref{sec:PCA2D}, PCA2D is used to analyze the peak intensity maps of 15 isolated CH$_3$OH lines. To identify meaningful PCs, we examined the variation of $\lambda_i$ with the order of PCs \citep{PCABook,Yun23}. As the order number ($i$) increases, noise dominates and $\lambda_i$ converges toward a small constant value, while the differences between the $\lambda_i$ values ($\lambda_{i-1} - \lambda_i$) approach zero (Fig.~\ref{fig:scree}). Meaningful PCs can, therefore, be identified as those with $\lambda_i$ that significantly deviates from this convergence. Figure~\ref{fig:scree} shows that $\lambda_i$ becomes nearly constant beyond PC5. Thus, four of the 15 PCs capture significant variation among the CH$_3$OH lines and are considered meaningful components in the PCA2D result.

\end{appendix}

\end{document}